\documentclass[conference]{IEEEtran}
\IEEEoverridecommandlockouts

\usepackage{amsmath}

\usepackage{cite}
\usepackage{amsmath,amssymb,amsfonts}
\usepackage{algorithmic}
\usepackage[]{graphicx}
\graphicspath{{Figures/}}
\usepackage{caption}
\usepackage{subfig}
\usepackage{lipsum}
\usepackage{balance}
\usepackage{textcomp}
\usepackage{xcolor}
\usepackage{multirow}
\usepackage{color}
\usepackage{algorithm} 
\usepackage{algorithmic}

\def\BibTeX{{\rm B\kern-.05em{\sc i\kern-.025em b}\kern-.08em
    T\kern-.1667em\lower.7ex\hbox{E}\kern-.125emX}}

\begin{document}

\title{Solar Irradiance Forecasting Using \\Triple Exponential Smoothing}

\author{
Soumyabrata~Dev$^{\dagger}$$^{1}$,
Tarek~AlSkaif$^{\dagger}$$^{2}$,
Murhaf~Hossari$^{1}$,
Radu~Godina$^{3}$,
Atse~Louwen$^{2}$,
and~Wilfried~van~Sark$^{2}$
\\
$^{1}$The ADAPT SFI Research Centre, Trinity College Dublin, Dublin, Ireland \\
$^{2}$~Copernicus Institute of Sustainable Development, Utrecht University, Utrecht, The Netherlands\\
$^{3}$~Department of Electromechanical Engineering (C-MAST), University of Beira Interior, Covilh\~a, Portugal
\thanks{$^{\dagger}$~Authors contributed equally.}
\thanks{
This work is supported by the the Joint Programming Initiative (JPI) Urban Europe project: “PARticipatory platform for sustainable ENergy managemenT (PARENT)” and the Netherlands Science Foundation (NWO).
}
\thanks{
Send correspondence to T. AlSkaif, E-mail: t.a.alskaif@uu.nl
}
}

\maketitle
\IEEEpubid
{\begin{minipage}{\textwidth}\ \\[12pt] \centering
  978-1-5386-5326-5/18/\$31.00~\copyright~2018 IEEE
\end{minipage}} 

%
\begin{abstract}
Owing to the growing concern of global warming and over-dependence on fossil fuels, there has been a huge interest in last years in the deployment of Photovoltaic (PV) systems for generating electricity. The output power of a PV array greatly depends, among other parameters, on solar irradiation. However, solar irradiation has an intermittent nature and suffers from rapid fluctuations. This creates challenges when integrating PV systems in the electricity grid and calls for accurate forecasting methods of solar irradiance. In this paper, we propose a triple exponential-smoothing based forecasting methodology for intra-hour forecasting of the solar irradiance at future lead times. We use time-series data of measured solar irradiance, together with clear-sky solar irradiance, to forecast solar irradiance up-to a period of $20$ minutes. The numerical evaluation is performed using $1$ year of weather data, collected by a PV outdoor test facility on the roof of an office building in Utrecht University, Utrecht, The Netherlands. We benchmark our proposed approach with two other common forecasting approaches: persistence forecasting and average forecasting. Results show that the TES method has a better forecasting performance as it can capture the rapid
fluctuations of solar irradiance with a fair degree of accuracy.

\end{abstract}

\begin{IEEEkeywords}
Photovoltaic, Solar energy analytics, Clear sky model, Intra-hour forecasting, Exponential smoothing.
\end{IEEEkeywords}

\section{Introduction}
\label{sec:1}
Due to the increasing concerns on greenhouse gases, the transition towards a more renewable energy system is becoming more visible. This trend has manifested itself by a sharp increase in the deployment of Photovoltaic (PV) systems for generating electricity in both standalone and grid-connected residential and large-scale systems due to their economical and environmental benefits~\cite{reinders2017photovoltaic}. 

Although PV systems generally operate reliably and over lifetimes of 20+ years, the integration of PV in electricity grids is hampered by their intermittent power output. The power output of PV systems can be highly variable due to the variability of solar global irradiation and the effect of different meteorological variables, such as temperature, humidity level, air pressure and cloud cover. This variability can create serious issues for the electricity grid needs, such as power system control, grid integration and power planning, especially at high penetration levels of PV~\cite{raza2016recent, van2018techno}.

In order to ensure reliable and stable operation of the electricity grid, it is important to accurately forecast the solar irradiance and predict its variability over time. As a result, the field solar irradiance forecasting has received an increasing attention among researchers over the past decade~\cite{raza2016recent, antonanzas2016review, kleissl2013solar}. The literature is rich with various methods for forecasting solar irradiance. A comprehensive review on the recent advances in PV power forecasting are presented in~\cite{raza2016recent, antonanzas2016review}. These papers provide overviews on: i) the factors that affect the PV power forecasting, ii) the different methods of forecasting; including persistence methods, physical models and machine learning models, and iii) forecasting performance evaluation metrics.  

Solar irradiance forecasting can be classified into different categories, depending on the forecasting horizon, ranging from intra-hour or intra-day forecast to days or weeks ahead forecast. 
Very short term solar irradiance forecasting (i.e., intra-hour forecast) is very important to assure grid stability, power quality, to reformulate bids in sub-hourly markets and to correctly plan reserve capacity and demand response, especially in locations or applications where high solar penetration is present \cite{antonanzas2016review}. In intra-hour PV solar irradiance forecasting, the forecasting horizon ranges from a few seconds to 1 hour and it has been addressed using different forecasting methods in~\cite{jang2016solar, elsinga2017short, chow2011intra, dev2016estimation, marquez2012comparison, cros2013clear, 8260177}. The work in~\cite{jang2016solar} uses a forecasting model based on satellite images and a support vector machines (SVM) learning scheme. A PV power forecasting in a network of neighboring PV systems is proposed in~\cite{elsinga2017short}. The work is based on the cross-correlation time lag between clear-sky index of those PV systems that are influenced by the same cloud. An estimation of solar irradiance using ground-based whole sky imagers is proposed in~\cite{chow2011intra, dev2016estimation}. An analysis of different clear sky models for the evaluation of PV output forecast is proposed in~\cite{marquez2012comparison, cros2013clear}. A Markov chain-based forecasting approach is used in~\cite{8260177} for calculating the expected value of future PV output power.

In this paper, we address the problem of intra-hour solar irradiance forecasting by asking the following question: can we predict the very short term future solar irradiance values, using historical measured values of solar irradiance? To answer this question, we propose a new method of solar irradiance forecasting, using Triple Exponential Smoothing (TES), that exploits the seasonality of the measured solar irradiance to predict future values up-to a period of $20$ minutes. We use time-series data of measured- and theoretical clear-sky- solar irradiance values for the prediction of future solar irradiance values. To the best of our knowledge, this is the first time the seasonality of the measured solar irradiance values using TES is exploited for very short term solar irradiance forecasting. We benchmark our proposed approach with other methods and we obtain competitive forecasting results. 

The structure of the paper is organized as follows. Section~\ref{sec:2} describes the clear sky model that is used in this paper. The triple exponential smoothing method is described in Section~\ref{sec:2}. Section~\ref{sec:4} presents and discusses the numerical results. Finally, we conclude the paper and give pointers for future directions in Section~\ref{sec:5}.

\section{Clear Sky Model}
\label{sec:2}
The total solar irradiation falling on the earth's surface in the event of a clear sky day is popularly referred as the clear sky solar irradiance. There are several theoretical models that estimates the amount of clear sky irradiance for a particular location on the earth's surface. This is dependent on a number of factors, including the time of the day, azimuth- and elevation- angles of the sun, and other atmospheric conditions. In our previous work~\cite{dev2017study}, we have provided a comparative analysis of various clear-sky models for a particular experimental location in the tropical region of Singapore.

In this paper, we use the popular Bird's clear sky model, proposed by Bird and Hulstrom~\cite{bird1981simplified}. It estimates the total clear sky irradiance by using the results from radiative transfer codes. This model is widely used by solar analytics experts because it considers the various atmospheric conditions, including the amount of aerosols, ozone and water vapor. The clear sky radiation~\cite{annear2007comparison}, $\phi_s$ (W/m$^2$) is given by:

\begin{equation}
\begin{aligned}
\label{eq:bird}
\phi_s = \frac{(\phi_d + \phi_l)}{(1-R_gr_s)},
\end{aligned}
\end{equation}
where $\phi_d$ denotes the direct solar irradiance, $\phi_l$ is the scattered solar irradiance, $R_g$ is the water surface reflectivity, and $r_s$ is the atmospheric albedo. The direct solar irradiance $\phi_d$, also measured in W/m$^2$, is computed via:

\begin{equation}
\begin{aligned}
\label{eq:direct}
\phi_d = 0.9662\phi_{ext}T_AT_WT_{UM}T_0T_R,
\end{aligned}
\end{equation}
where the parameter $\phi_{ext}$ is the extraterrestrial solar irradiance, $T_A$ is the transmittance of aerosol absorptance and scattering, $T_W$ is the transmittance of water vapor, $T_{UM}$ is the transmittance of uniformly mixed gases, $T_0$ is the transmittance of ozone content, and $T_R$ is the transmittance of Rayleigh scattering. 

The scattered solar irradiance $\phi_l$, is also measured in W/m$^2$, and is given by:

\begin{equation}
\begin{aligned}
\label{eq:scatter}
\phi_l = 0.79\phi_{ext}T_{AA}T_WT_{UM}T_0\frac{0.5(1-T_R)+B_a(1-\frac{T_A}{T_{AA}})}{1-m_p+m_p^{1.02}},
\end{aligned}
\end{equation}
where the $T_{AA}$ is the transmittance of aerosol absorptance, $B_a$ is a dimensionless constant, and $m_p$ is the relative optical air mass. We use the python implementation of Bird's clear sky model for estimating the total received solar irradiance, on a clear sky day~\cite{witmer2015python}. This assists us in estimating the solar irradiance on a particular location of the earth, for a clear sky day.

However, in most of the days, the actual solar irradiance varies from this theoretical clear sky irradiance value. This is mainly because of the clouds and other atmospheric events in the atmosphere -- the actual solar irradiance differs from the clear sky irradiance value. We define the clearness index ($k$) as the ratio of measured solar irradiance ($\phi_{act}$) and the theoretical clear sky irradiance ($\phi_s$) data.  

\begin{equation}
\begin{aligned}
\label{eq:k-value}
k = \phi_{act}/\phi_s.
\end{aligned}
\end{equation}

We model this clearness index ($k$) based on the actual- and clear-sky- irradiance data. This provides us an estimate on the condition of the sky (clear sky or overcast) at a particular time instant.

\section{Triple Exponential Smoothing}
\label{sec:3}
In order to estimate the future values of solar irradiance, we use the Triple Exponential Smoothing (TES) method. In general, the exponential smoothing function is a straightforward method for smoothing time series data. It assigns exponentially decreasing weights over time~\cite{gardner1985exponential}. When dealing with high frequency signals, it is common to to use a TES function. It can be used three times to smooth a three-signals time series and remove high frequencies encountered~\cite{kalekar2004time}. 

Suppose we have a sequence of clear sky indices (i.e., time series data) $k_t$, for a cycle of seasonal change that has the length $L$ and in case we are starting at $t=0$. The TES function provides a best estimate of the future clearness index value at time $t+1$ ($k_{t+1}$) (i.e., the smoothed value) using the clearness index value at time $t$ as an input ($k_t$). The TES computes a line that follows the trend of the data (i.e., clear sky indices). Similarly, it calculates the seasonal indices which are used to weigh the corresponding values in the trend line by fitting the time point to the cycle of length $L$. 

For a given time t, we can represent the smoothed value of the clearness index by $s_t$. To represent for a given linear trend the sequence of estimates we use $b_t$. Those estimates would be the best estimates that are placed on the seasonal changes. Seasonal correction factors can be represented by $c_t$. For a give time $t \mod L$ in the cycle where we took the observations on, $c_t$ stands for the expected proportion of the trend that was predicted.

In order to intialize a set of seasonal factors, we need a minimum of $2L$ periods, which translates to two full seasons of historical data.

The TES method's output ($F_{t+m}$) is a predicted estimation for the value of $k$ at a time $t+m, m>0$. This predication is based on that data seen before and up to the time $t$. The TES with additive seasonality is given by the following formulas \cite{natrella2010nist}:

\begin{align}
s_0 &=k_0,\\
s_t &=\alpha (k_t -c_{t-L}) + (1-\alpha)(s_{t-1}+b_{t-1}),\\
b_t &=\beta (s_t - s_{t-1}) + (1-\beta)b_{t-1},\\
c_t &=\gamma (k_t - s_{t-1} -b_{t-1})  + (1-\gamma)c_{t-L},\\
F_{t+m}&=s_t+mb_t + c_{t-L+1+(m-1) \mod L},
\end{align}
where the data smoothing factor is $\alpha$, $0< \alpha <1$, the trend smoothing factor is $\beta$, $0< \beta <1$, and  the seasonal change smoothing factor is $\gamma$, $0< \gamma <1$.

\begin{align}b_{0}&={\frac {1}{L}}\left({\frac {k_{L+1}-k_{1}}{L}}+{\frac {k_{L+2}-k_{2}}{L}}+\ldots +{\frac {k_{L+L}-k_{L}}{L}}\right).\end{align}

Setting the initial estimates for the seasonal indices $c_i$ for $i = 1,2,\ldots,L$ is a bit more involved. If $N$ is the number of complete cycles present in our data, then:

\begin{align}
c_{i}&={\frac {1}{N}}\sum _{j=1}^{N}{\frac {k_{L(j-1)+i}}{A_{j}}}\quad \forall i&=1,2,\ldots ,L,
\end{align}
where in jth cycle of our data, $A_j$ stands for the average value of $x$ and it can be calculated as follows:

\begin{align}
A_{j}&={\frac {\sum _{i=1}^{L}k_{L(j-1)+i}}{L}}\quad \forall j&=1,2,\ldots ,N.
\end{align}

In summary, our proposed forecasting algorithm can be summarized in Algorithm~\ref{A:TESSolar}.

\begin{algorithm}[htb]
	\caption{TES based solar irradiance forecasting}
    \label{A:TESSolar}
	\begin{algorithmic}[1]
		\REQUIRE Time series data of measured solar irradiance, latitude and longitude of the region of interest.
		\STATE Ensure that the geographical co-ordinates of the place is accurate;
		\STATE Set the period of historical data (at least $2L$ time periods) for training the TES model;
        \STATE Compute the Bird's clear sky model for the considered historical time period;
		\STATE Generate the values of the clearness index $k$, for the same time period, as described in Section~\ref{sec:2};
        \STATE Train the TES model using the computed $k$ values of the historical time period;
        \STATE Estimate the future values of $k$ values for the considered lead time;
		\RETURN Estimated $k$ values, required in computing forecasted solar irradiance.
	\end{algorithmic}
\end{algorithm}

\section{Results and Discussion}
\label{sec:4}

\subsection{Data Collection}
\begin{figure}[htb]
\begin{center}
	\includegraphics[width=0.9\linewidth]{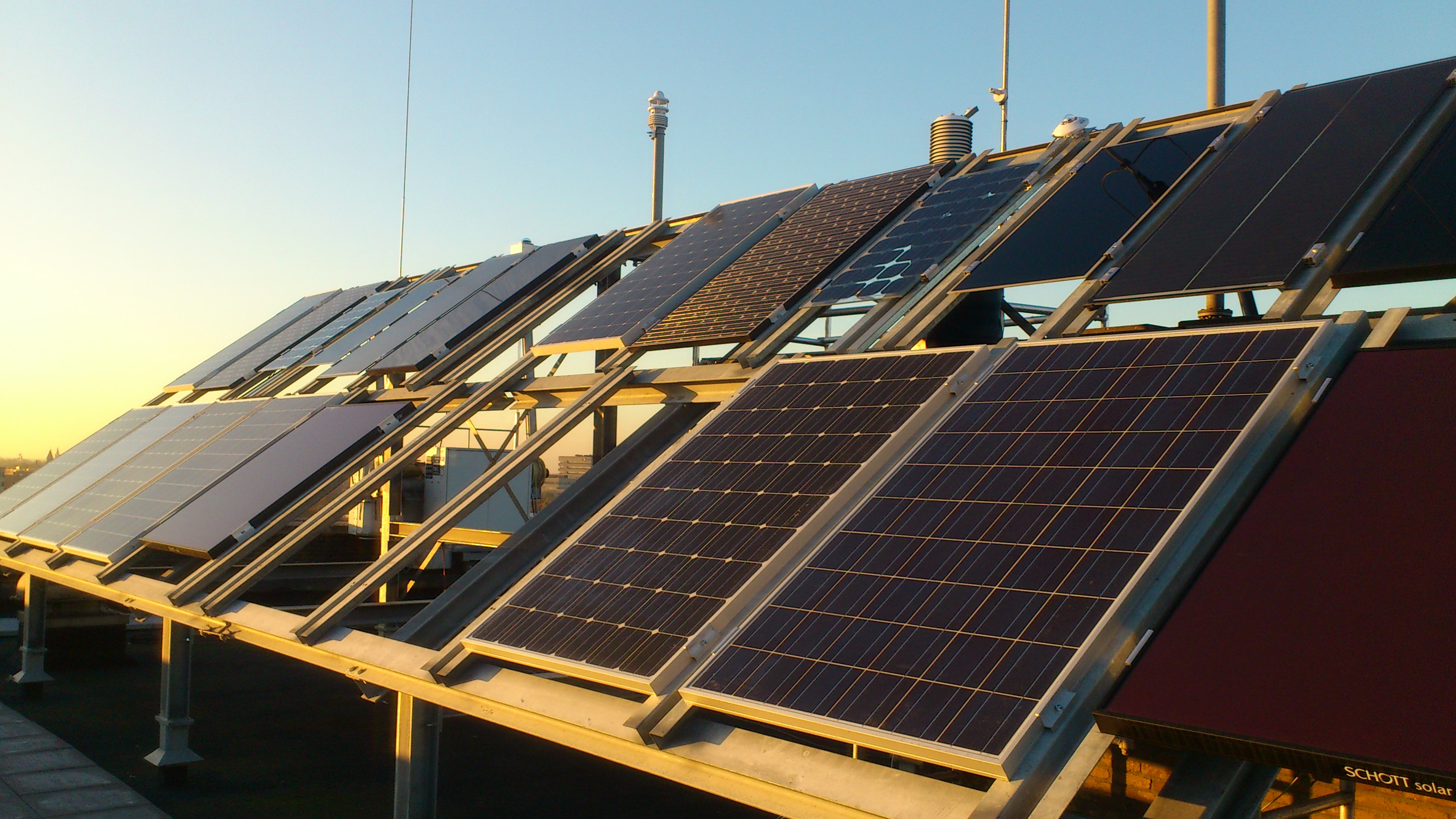}
	\caption{Photograph of the Utrecht Photovoltaic Outdoor Test facility (UPOT). Photo courtesy of Arjen de Waal. (Best viewed in color)}
	\label{fig:upot}
\end{center}
\end{figure}

The data used here is collected from a rooftop PV test facility at Utrecht University: the Utrecht Photovoltaic Outdoor Test facility (UPOT), shown in Figure~\ref{fig:upot}. 
The facility is located at Utrecht University's campus in the center of the Netherlands on top of an eight story building. It is equipped with $24$ PV modules of different brands and technologies, of which IV curves and back-of-module temperature are measured at a time-resolution of $2$ minutes. The test facility has an array of solar irradiance sensors measuring direct normal irradiance (DNI) and diffuse horizontal irradiance (DHI), global horizontal irradiance (GHI), global in-plane irradiance, and spectral irradiance. Furthermore, air temperature, wind speed and other meteorological data are measured. These measurements are taken at a time-resolution of $30$ seconds. More details about UPOT test facility can be found in ~\cite{van2012upot}.

\begin{figure}[htb]
\begin{center}
\subfloat[Clear-sky day]{%
\includegraphics[width=0.45\textwidth]{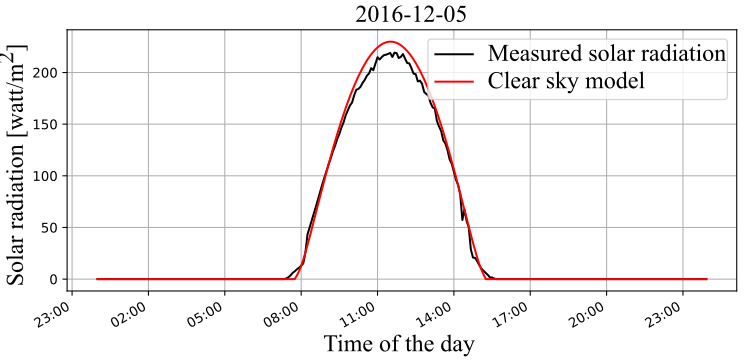}}\\
\subfloat[Overcast day]{%
\includegraphics[width=0.45\textwidth]{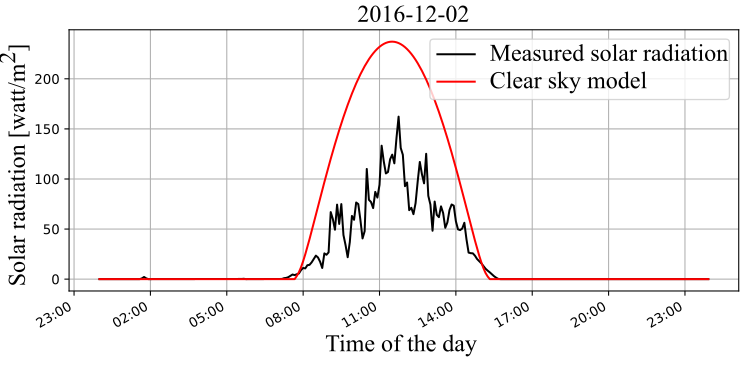}}
\caption{Illustration showing the measured solar irradiance and theoretical clear-sky irradiance, on the event of (a) clear-sky day and (b) overcast day. (Best viewed in color)
\label{fig:clear-sky}}
\end{center}
\end{figure}

In this paper, we use the measured solar irradiance data for $1$ year, collected at the UPOT site. We use the Bird's clear sky model to compute the theoretical clear sky irradiance for a particular date. Figure~\ref{fig:clear-sky}(a) illustrates the measured solar irradiance and the Bird's clear sky irradiance for a sample clear sky data. We observe that the theoretical model closely follows the actual solar irradiance measured at the site. Interestingly, on a non-clear sky day as shown in Fig.~\ref{fig:clear-sky}(b), there are rapid fluctuations in the measured solar irradiance. These fluctuations occur because of the clouds and other atmospheric events that impact the received solar irradiance.

\begin{figure}[htb]
\centering
\includegraphics[width=0.3\textwidth]{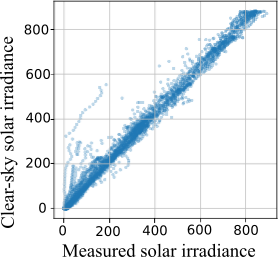}
\caption{Scatter plot between measured solar irradiance and computed clear sky solar irradiance, for all the clear sky days in 2015. The Pearson degree of correlation is $0.99$.}
\label{fig:scatter}
\end{figure}

In order to evaluate the performance of Bird's clear sky model at our Utrecht solar irradiance data, we analyze the relationship between measured- and clear sky- solar irradiance data. We manually list all the clear sky days in the year of $2015$ from the solar irradiance measurements. We compute the corresponding clear-sky irradiance values for all the available solar irradiance recordings of clear sky days. We observe that there are a total of $22$ clear sky days in the year $2015$ at Utrecht. Figure~\ref{fig:scatter} shows the scatter plot between the measured- and Bird's clear sky- irradiance for all the $22$ clear sky days. We compute the Pearson degree of correlation between the two values and obtain a high value of $0.99$. This high degree of correlation indicate that the Bird's clear sky can model the solar irradiance value (on a clear day) with a high degree of accuracy.

However, in most of the days, the measured solar irradiance fluctuates from the theoretical clear sky data (cf.\ Fig.~\ref{fig:clear-sky}b). These fluctuations are detrimental in the context of solar energy analytics, as it is very difficult to track and predict the fluctuations. In this paper, we attempt to forecast these fluctuations in the future using the method proposed in Section~\ref{sec:3}, based on historical data of measured solar irradiance and theoretical clear sky model.

\subsection{Statistical Results}

\begin{figure}[htb]
\centering
\includegraphics[width=0.45\textwidth]{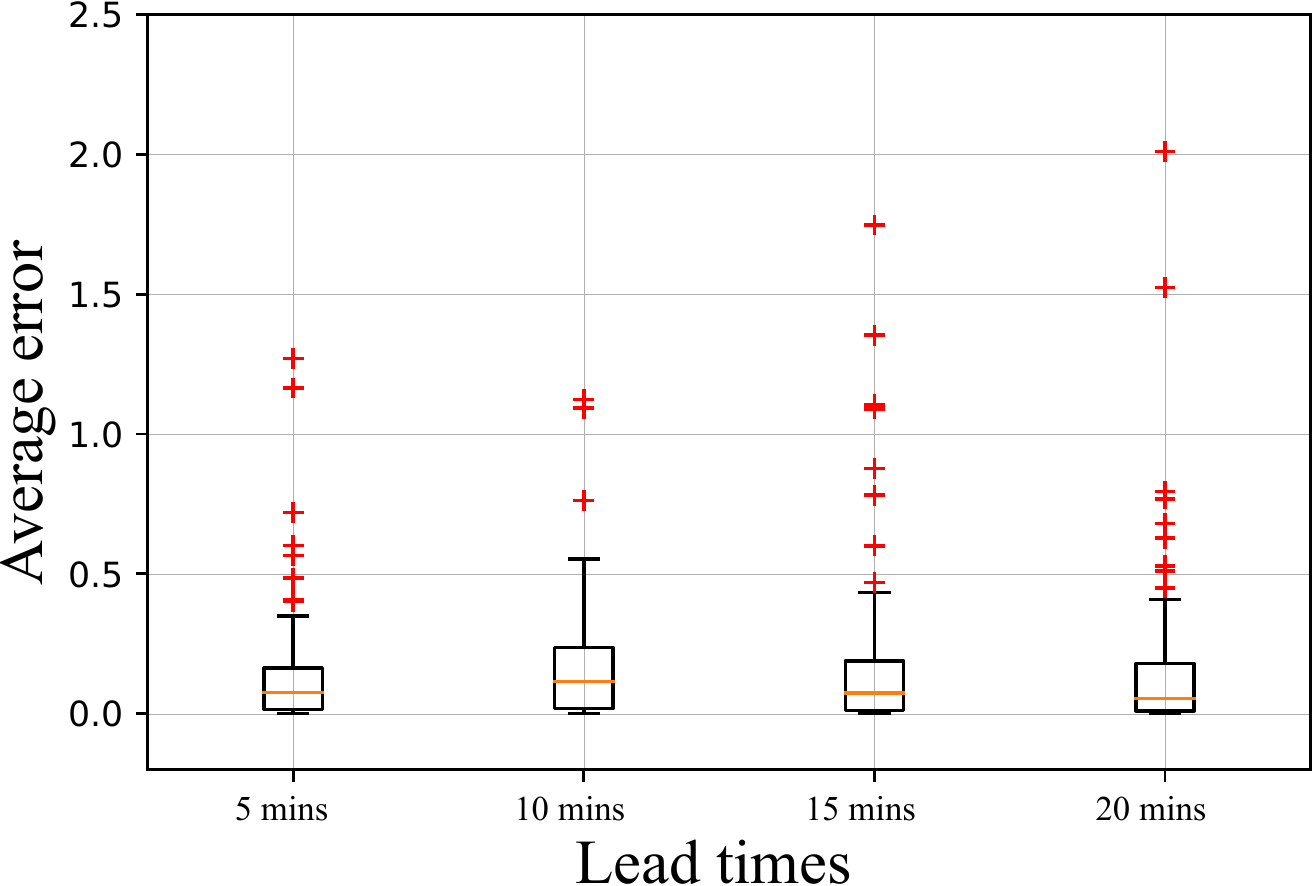}
\caption{Trend of average error (i.e., MAE) in estimating $k$ for various lead times. We perform $150$ experiments for a particular lead time.}
\label{fig:stat-results}
\end{figure}

In this paper, we consider the meteorological measurements for the entire year of $2015$, as recorded by the UPOT test facility. All measurements are resampled in intervals of $5$ minutes. Therefore, we set the seasonal period parameter in Holt-Winters as $L=288$, as repeatability occurs after every $288$ observations (i.e., one day). In this section, we perform a statistical analysis of our proposed solar forecasting methodology. We use a total of $1728$ observations (or $6L$ periods) for training our model, and attempt to predict the clearness index values up to a period of $20$ minutes. We compute the Mean Absolute Error (MAE) between the predicted- and the actual- $k$ values to evaluate the performance of the model. These samples of timestamps are sampled randomly over the period of all the observations in the year of $2015$. We perform $150$ such experiments, sampled at random time stamps of the year. Figure~\ref{fig:stat-results} shows the box-plot of the error values for the different lead times. The distributions of the errors for different lead times show a clear trend -- the error increases gradually with increasing lead times. 

\subsection{Benchmarking}
We compare the performance of our proposed approach using TES with other two common forecasting approaches, viz.\ persistence forecasting and average forecasting. The persistence forecasting method assumes that the conditions for the forecast do not change (especially for shorter lead times). This method works well in conditions when weather maps change very slowly with time. We also compare our approach with the average forecasting method. Such method assumes that the average value remains the same while the response value fluctuates randomly for the shorter time period. Such technique assumes that the forecast value is the average of the previous observed recordings. 

\begin{table}[htb]
\normalsize
\centering
\caption{Benchmarking of several forecasting methods.}
\label{tab:diff-methods}
\begin{tabular}{l|c}
\hline
\textbf{Methods} & \textbf{Average Error ($100$ experiments)} \\ \hline
Persistence      & 0.41                   \\ 
Average          & 0.36                   \\ 
Proposed TES       & 0.13                   \\ \hline
\end{tabular}
\end{table}

\begin{figure*}[htb]
\begin{center}
\subfloat[Prediction for short lead times]{%
\includegraphics[width=0.9\textwidth]{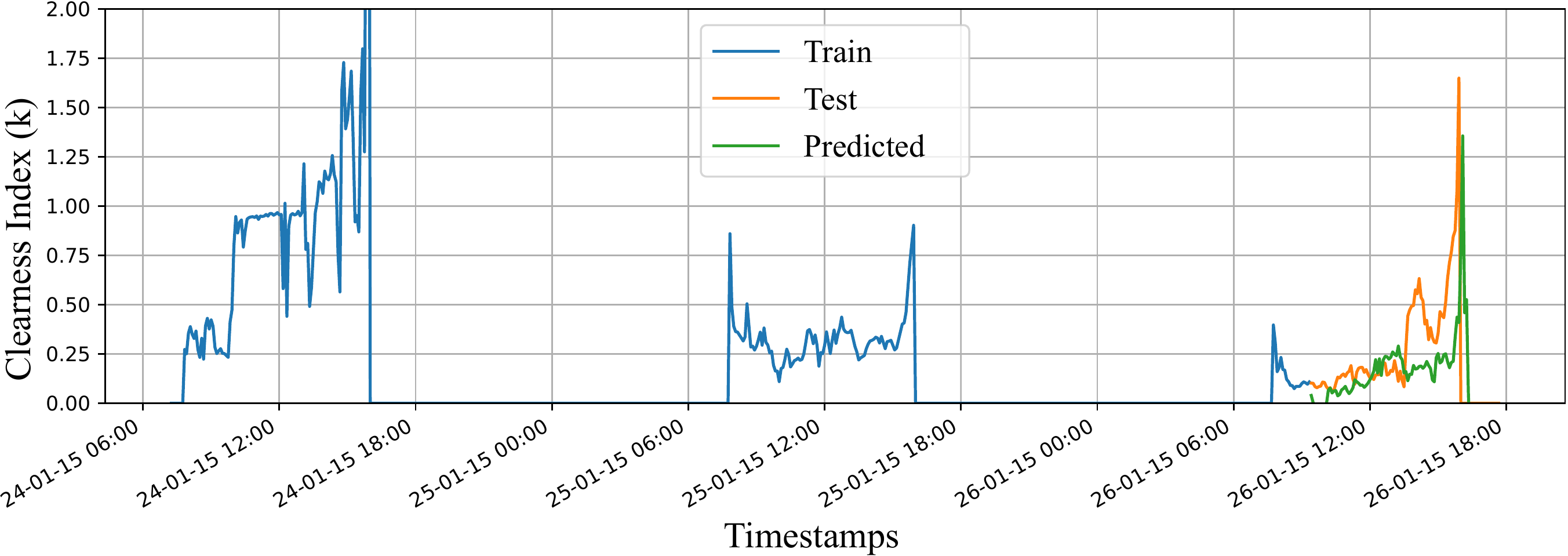}}\\
\subfloat[Prediction for longer lead times]{%
\includegraphics[width=0.9\textwidth]{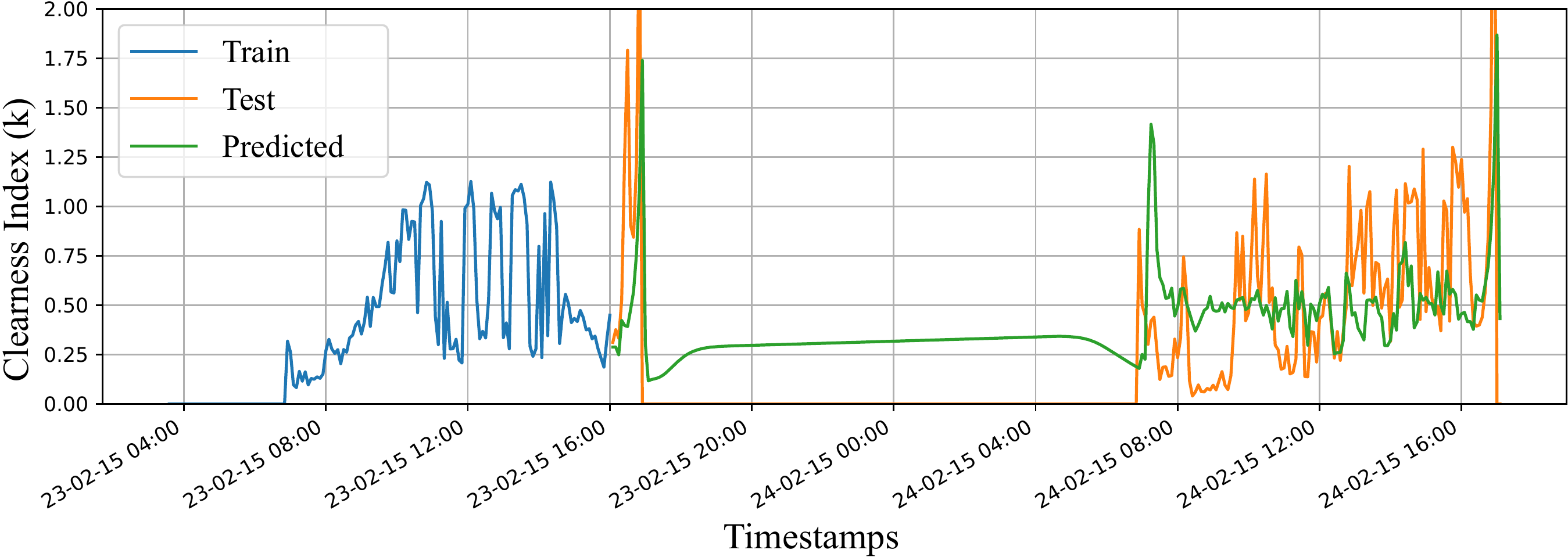}}
\caption{Prediction of future $k$ values using our proposed approach. We observe that the trend of the predicted $k$ values closely follows the trend of the actual $k$ values. (Best viewed in color)
\label{fig:sample-results}}
\end{center}
\end{figure*}

We compare the performance of these methods using our Utrecht solar irradiance data. We compute the forecast values of clearness index $k$, upto lead times of $20$ minutes, at intervals of $5$ minutes. We use historical data of $1152$ observations (or $4L$ periods) for the purpose of forecasting. The entire dataset is sampled at random time stamps to perform the forecasting accuracy. We perform a total of $100$ experiments to remove sampling bias. Table~\ref{tab:diff-methods} discusses the results. As expected, the performance of the persistence method is poor. The average method has a better performance, as it can capture the trend of the $k$ values. However, our proposed approach using TES has the best performance, as it can also capture the rapid fluctuations of $k$ values and predict it with a good degree of accuracy.

\subsection{Time Series Analysis}


As a final comparison, we visualize the performance of our proposed approach in the form of clearness index time series. Figure~\ref{fig:sample-results} shows the time-series results for two distinct time series. We observe that for a shorter lead time, as shown in Fig.~\ref{fig:sample-results}(a), the predicted clearness index values has the similar trend as the actual index values. This is an interesting observation, as our proposed approach can capture the rapid fluctuations of solar irradiance with a fair degree of accuracy. Similarly, for a longer lead time (cf.\ Fig.~\ref{fig:sample-results}b), the trend of the $k$ values is captured well even for the next day timestamps. However, the predicted $k$ values missed the very short-term variations. This is promising for us, as our proposed approach can provide a reliable methodology for estimating the future solar irradiance.

\section{Conclusions and Future Work}
\label{sec:5}
In this paper, we have proposed a methodology for intra-hour forecasting of solar irradiance, based on the historical data of the clearness index. Our method is based on a triple exponential smoothing function that captures the seasonality of our dataset with high degree of accuracy. It takes into account both the trend and seasonality to forecast future intra-hour solar irradiance values. 

Our future work involves the validation of our proposed approach using more than $1$ year of meteorological data. We also plan to use images captured by ground-based sky cameras~\cite{dev2014wahrsis,dev2015design} to further enhance the prediction results. Our initial results in \cite{dev2016estimation} indicate that image-based features obtained from sky cameras can accurately estimate the solar irradiance falling on earth's surface. Unlike point-measurement devices, sky cameras provide additional information about cloud movement and coverage. This additional information will equip us with a multi-modal approach in predicting future solar irradiance with a high degree of accuracy.

\balance

\end{document}